\documentclass[aps,pra,preprint,showpacs,showkeys]{revtex4}
\usepackage{amssymb,bm}
\usepackage{graphicx}
\usepackage{amsmath}
\usepackage{epstopdf}
\usepackage{float}
\allowdisplaybreaks

\begin{document}

\title{Integrated cross sections of high-energy $e^+e^-$ electroproduction  by an electron in an atomic field}

\author{P. A. Krachkov}\email{peter_phys@mail.ru}
\author{A. I. Milstein}\email{A.I.Milstein@inp.nsk.su}
\affiliation{Budker Institute of Nuclear Physics, 630090 Novosibirsk, Russia}

\date{\today}

\begin{abstract}

The integrated cross sections of high-energy $e^+e^-$ electroproduction  by an electron in an atomic field is studied.  Importance of various contributions to these cross sections is discussed.
It is shown that the Coulomb corrections are very important both for the differential cross section and for the integrated cross sections even for moderate values of the atomic charge number.
This effect is mainly related to the  interaction of the produced $e^+e^-$ pair with the atomic field. For the  cross section differential over the electron transverse  momentum the account for the interference of the amplitudes  and the contribution of  virtual bremsstrahlung is noticeable.  The Coulomb corrections to  scattering is larger than these two effects but essentially smaller than the Coulomb corrections to the amplitude of  pair photoproduction by a virtual photon. However,  in the  cross section differential over the positron transverse  momentum,  there is a strong suppression of the effect of the Coulomb corrections to scattering.
\end{abstract}



\maketitle

\section{Introduction}
 A process of $e^+e^-$ pair production by a high-energy electron in the atomic field is interesting both from experimental and theoretical points of view. It is important to know the cross section of this process with high accuracy at data analysis in detectors.  Besides, this process gives the substantial contribution to a background at precision experiments devoted to search of new physics. From the theoretical point of view, the cross section of electroproduction in the field of heavy atoms reveals very interesting properties  of the Coulomb corrections, which are the difference between the cross section exact in the parameters of the field and that calculated in the lowest  order of the perturbation theory (the Born approximation).

The cross sections in the Born approximation, both differential and integrated, have been discussed in numerous papers
\cite{Bhabha2,Racah37,BKW54,MUT56,Johnson65,Brodsky66,BjCh67,Henry67,Homma74}. The Coulomb corrections to the differential cross section of high-energy electroproduction by an ultra-relativistic electron in the atomic field have been obtained only recently  in our paper  \cite{KM2016}. In that  paper it is shown  that the Coulomb corrections significantly modify the differential cross section of the process as compared  with the Born result.  It turns out that both effects, the  exact account for the interaction of incoming and outgoing electrons with the atomic field and the exact account for the interaction of the produced pair with the atomic field, are very important for the value of the differential cross section. On the other hand, the are many papers devoted to the calculation of  $e^+e^-$ electroproduction  by a heavy particles (muons or nuclei) in an atomic field  \cite{Nikishov82,IKSS1998,SW98,McL98,Gre99,LM2000}. It that papers, the interaction of a heavy particle and the atomic field have been neglected. In our recent paper \cite{KM2017} it has been shown that the cross section, differential over the angles of a heavy outgoing particle, changes significantly due to the exact account for the interaction of a heavy particle with the atomic field.  However, the cross section integrated over these angles is not affected by this interaction. Such unusual properties of the cross section of electroproduction by a heavy particle  stimulated us to perform the detailed investigation of the integrated cross section of the electroproduction by the ultra-relativistic electron.

In the present paper we investigate in detail  the integrated cross section, using the analytical result for the matrix element of the process obtained in our  paper \cite{KM2016} with the exact account for the interaction of all charged particles with the atomic field.  Our goal is to understand the relative importance of various contributions to the integrated cross section under consideration.

\section{General discussion}\label{general}

\begin{figure}[h]
\centering
\includegraphics[width=1.\linewidth]{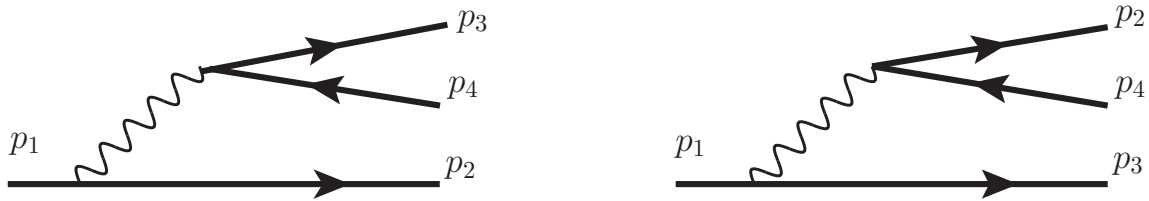}
\caption{Diagrams $T$ (left) and  $\widetilde{T}$ (right) for the contributions to the amplitude ${\cal T}$ of the process $e^-Z\to e^- e^+e^-Z$. Wavy line denotes the photon propagator, straight lines denote the wave functions in the atomic field.}
\label{fig:diagrams}
\end{figure}

The differential cross section of high-energy electroproduction by an unpolarized electron  in the atomic  field   reads
\begin{equation}\label{eq:cs}
d\sigma=\frac{\alpha^2}{(2\pi)^8}\,d\varepsilon_3d\varepsilon_4\,d\bm p_{2\perp}\,d\bm p_{3\perp}d\bm p_{4\perp}\,\frac{1}{2}\sum_{\mu_i=\pm1}|{\cal T}_{\mu_1\mu_2\mu_3\mu_4}|^{2}\,,
\end{equation}
where $\bm p_1$ is the momentum of the  initial electron,  $\bm p_2$ and $\bm p_3$ are the final electron  momenta, $\bm p_4$ is the positron momentum, $\mu_i=\pm 1$ corresponds to the helicity of the particle with the momentum $\bm p_i$, $\bar\mu_i=-\mu_i$,  $\varepsilon_{1}=\varepsilon_{2}+\varepsilon_{3}+\varepsilon_{4}$ is the  energy of the  incoming electron,   $\varepsilon_{i}=\sqrt{{p}_{i}^2+m^2}$, $m$ is the electron mass, and $\alpha$ is the fine-structure constant,  $\hbar=c=1$.  In Eq.~\eqref{eq:cs} the notation  $\bm X_\perp=\bm X-(\bm X\cdot\bm \nu)\bm\nu$ for any vector $\bm X$ is used, $\bm\nu=\bm p_1/p_1$.
We have
\begin{equation}\label{TTT}
{\cal T}_{\mu_1\mu_2\mu_3\mu_4}=T_{\mu_1\mu_2\mu_3\mu_4}-\widetilde{T}_{\mu_1\mu_2\mu_3\mu_4}\,,
\quad \widetilde{T}_{\mu_1\mu_2\mu_3\mu_4}=T_{\mu_1\mu_3\mu_2\mu_4}(\bm p_2\leftrightarrow \bm p_3)\,,
\end{equation}
where the contributions $T$ and $\widetilde{T}$   correspond, respectively, to the left and right diagrams in Fig.~\ref{fig:diagrams}.
The amplitude $T$ has been derived in Ref.~\cite{KM2016} by means of the quasiclassical approximation \cite{KLM2016}. Its explicit form is given in  Appendix with one modification. Namely, we have introduced  the parameter $\lambda$ which is equal to unity if the  interaction of electrons, having  the momenta $\bm p_1$, $\bm p_2$  in the term $T$ and $\bm p_1$ , $\bm p_3$  in the term $\widetilde T$,  with the atomic field is taken into account. The parameter $\lambda$ equals to zero, if one neglects this interaction. Insertion of this parameter allows us to investigate the importance of various contributions to the cross section.

 First of all we note that the term $T$ is a sum of two contributions, see Appendix,
$$T=T^{(0)}+T^{(1)}\,,$$
where $T^{(0)}$ is the contribution to the amplitude in which the produced $e^+e^-$ pair does not interact with the atomic field, while the contribution $T^{(1)}$ contains such interaction.
In other words, the term $T^{(0)}$ corresponds to bremsstrahlung of the virtual photon decaying into a free  $e^+e^-$ pair. In the contribution $T^{(1)}$,  electrons with the momenta $\bm p_1$ and
$\bm p_2$ may interact or not interact with the atomic field. The latter contribution is given by the amplitude $T^{(1)}$ at $\lambda=0$. Below we refer  to the result of account for such  interaction  in the term $T^{(1)}$ as  the Coulomb corrections to scattering. Note that the contribution  $T^{(0)}$ at $\lambda=0$ vanishes.

In the present work we are going to elucidate  the following points:  the relative contribution of the term  $T^{(0)}$ to the cross section, an importance of the Coulomb corrections to scattering,  an importance of the interference between the amplitudes $T$ and $\widetilde{T}$ in the cross section.

\begin{figure}
\centering
\includegraphics[width=0.5\linewidth]{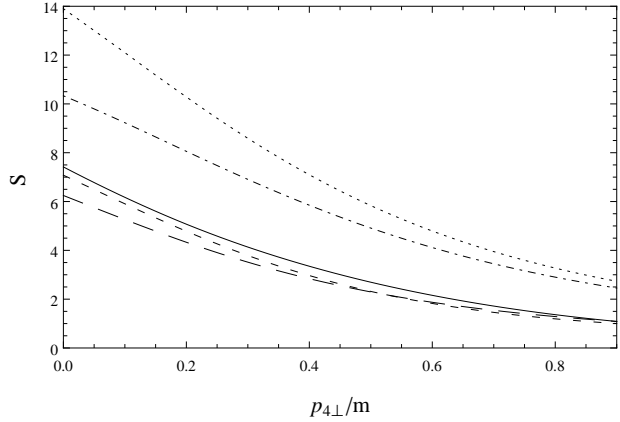}
\caption{The quantity $S$, see Eq. \eqref{S}, as the function of $p_{4\perp}/m$ for $Z=79$, $\varepsilon_1=100m$, $\varepsilon_2/\varepsilon_1=0.28$, $\varepsilon_3/\varepsilon_1=0.42$, $\varepsilon_4/\varepsilon_1=0.3$, $p_{2\perp}=1.3 m$, $p_{3\perp}=0.5 m$, $\bm p_{3\perp}$  parallel to $\bm p_{4\perp}$, and the angle between $\bm p_{2\perp}$ and $\bm p_{4\perp}$ being $\pi/2$; solid curve is the exact result,  dotted curve is the Born result, dash-dotted  curve is that obtained without account for the interference between $T$ and $\widetilde{T}$, the result for $\lambda=0$ is given by long-dashed curve, and the dashed curve corresponds to the result obtained by neglecting the contribution $T^{(0)}$ and $\widetilde{T}^{(0)}$.}
\label{dif}
\end{figure}

We begin our analysis with the case of the differential cross section. Let us consider the quantity $S$,
 \begin{equation}\label{S}
S=\sum_{\mu_i=\pm1}\Bigg|\frac{\varepsilon_1 m^4 {\cal T}_{\mu_1\mu_2\mu_3\mu_4}}{\eta (2\pi)^2}\Bigg|^2 \,,
\end{equation}
where $\eta=Z\alpha$ and $Z$ is the atomic charge number. In Fig.~\ref{dif}   the dependence of  $S$
 on the positron transverse momentum $p_{4\perp}$ is shown for gold ($Z=79$) at some values of $\varepsilon_i$, $\bm p_{2\perp}$, and  $\bm p_{3\perp}$.  Solid curve is the exact result, long-dashed curve corresponds to $\lambda=0$,  dashed curve is the result obtained without account for the contributions $T^{(0)}$ and $\widetilde{T}^{(0)}$, dash-dotted curve is the result obtained without account for the interference between $T$ and $\widetilde{T}$, and dotted curve is the Born result (in the Born approximation $S$ is independent of $\eta$). One can see for the case  considered, that the Born result differs significantly from the exact one, and account for the interference is also very important.  The contributions $T^{(0)}$ and $\widetilde{T}^{(0)}$ are  noticeable but not large, and the Coulomb corrections to the contributions $T^{(1)}$ and $\widetilde{T}^{(1)}$ are essential.
The effect of screening for the values of the parameters considered in Fig.~\ref{dif} is unimportant.  Note that  relative importance of different effects under discussions for the differential cross section strongly depends on the values of  $\bm p_{i}$. However, in all cases a deviation of the Born result from the exact one is substantial even for moderate values of $Z$.

\begin{figure}[H]
	\centering
	\includegraphics[width=0.45\linewidth]{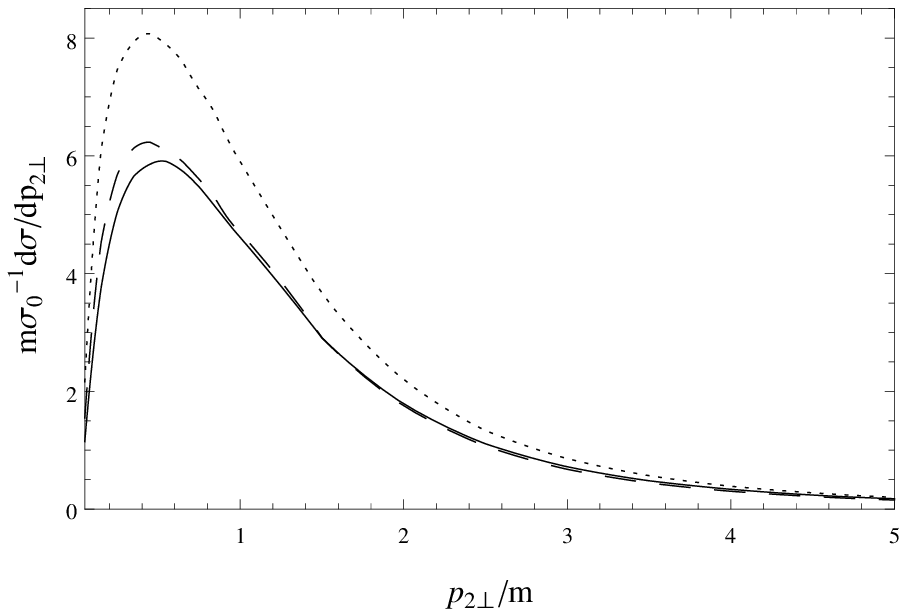}
	\includegraphics[width=0.45\linewidth]{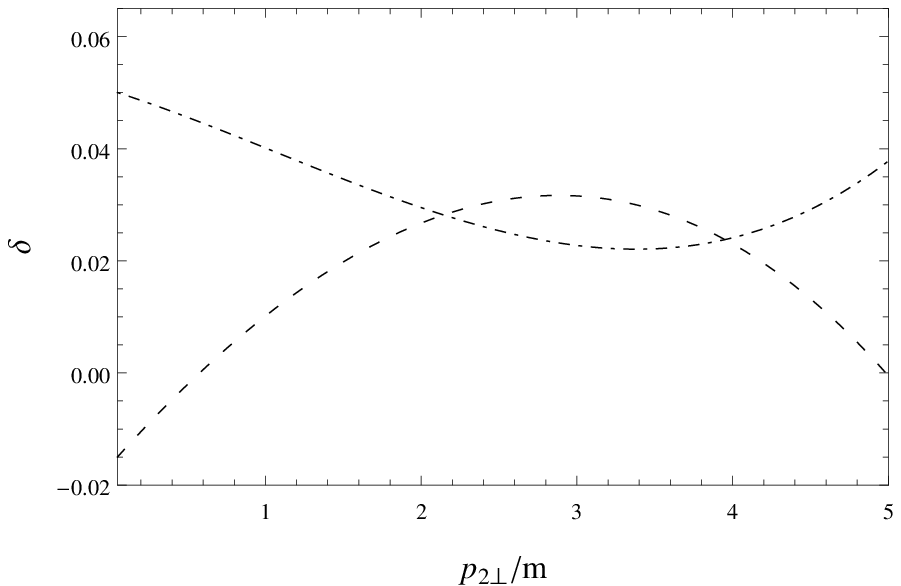}
	\caption{Left panel: the dependence of $d\sigma/dp_{2\perp}$ on $p_{2\perp}/m$ in units  $\sigma_0/m=\alpha^2\eta^2/m^3$ for  $Z=79$, $\varepsilon_1/m=100$; solid curve  is the exact result, dotted curve is the Born result, and long-dashed curve corresponds to $\lambda=0$. Right panel: the  quantity $\delta$ as the function of $p_{2\perp}/m$, where $\delta$ is  the deviation of the approximate result for $d\sigma/dp_{2\perp}$ from the exact one in units of the exact cross section. Dash-dotted curve is obtained without account for the interference between  $T$ and $\widetilde{T}$, dashed curve is obtained without contributions of $T^{(0)}$ and $\widetilde{T}^{(0)}$.}
\label{dif2}
\end{figure}

Let us consider the cross sections   $d\sigma/dp_{2\perp}$, i.e., the cross sections  differential over the electron transverse momentum $p_{2\perp}$.   This cross section for $Z=79$ and $\varepsilon_1=100 m$ is shown in the left panel in Fig.~\ref{dif2}. In this picture solid curve  is the exact result, dotted curve is the Born result, and long-dashed curve corresponds to $\lambda=0$.
It is seen that the exact result significantly differs from the Born one, and  account for the Coulomb corrections to scattering is also essential. An importance of account for the interference between  $T$ and $\widetilde{T}$, as well as account for the contributions of $T^{(0)}$ and $\widetilde{T}^{(0)}$, is demonstrated by the right  panel in Fig.~\ref{dif2}. In this picture the quantity $\delta$, which is  the deviation of the approximate result for $d\sigma/dp_{2\perp}$ from the exact one in units of the exact cross section, is shown. Dash-dotted curve is obtained without account for the interference between  $T$ and $\widetilde{T}$, dashed curve is obtained without contributions of $T^{(0)}$ and $\widetilde{T}^{(0)}$. It seen that both effects are noticeable.

Our  results are obtained under the condition $\varepsilon_i\gg m$, and  a question on  the limits of integration over energies appears at the numerical calculations of $d\sigma/dp_{2\perp}$. We have examined this question and found that the variation of the limits of integration in the vicinity of the threshold   changes only slightly the result of integration.
In any case, such a variation does not change the interplay of various contributions to the cross sections, and we present the results obtained by the integration over all kinematical region allowed.

\begin{figure}[H]
	\centering
	\includegraphics[width=0.45\linewidth]{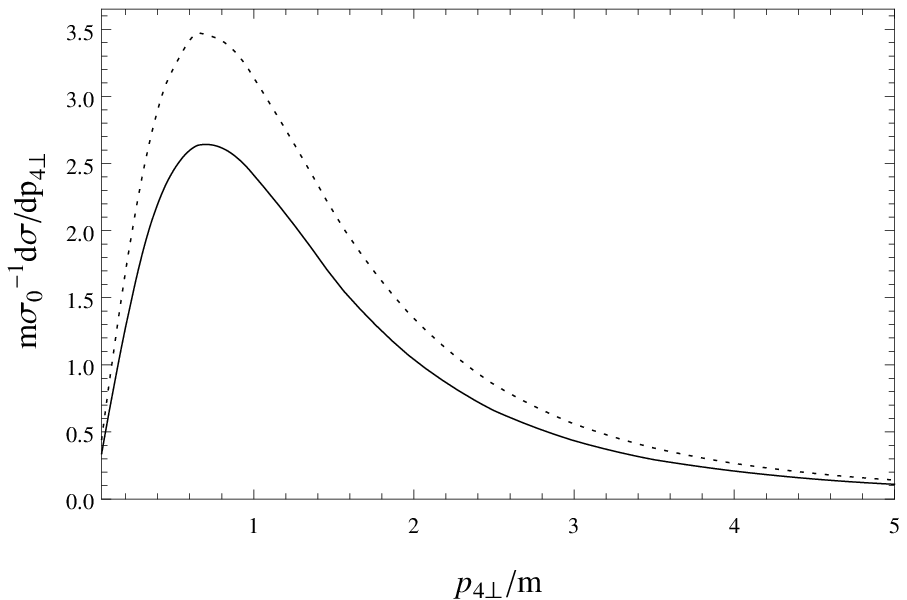}
	\includegraphics[width=0.45\linewidth]{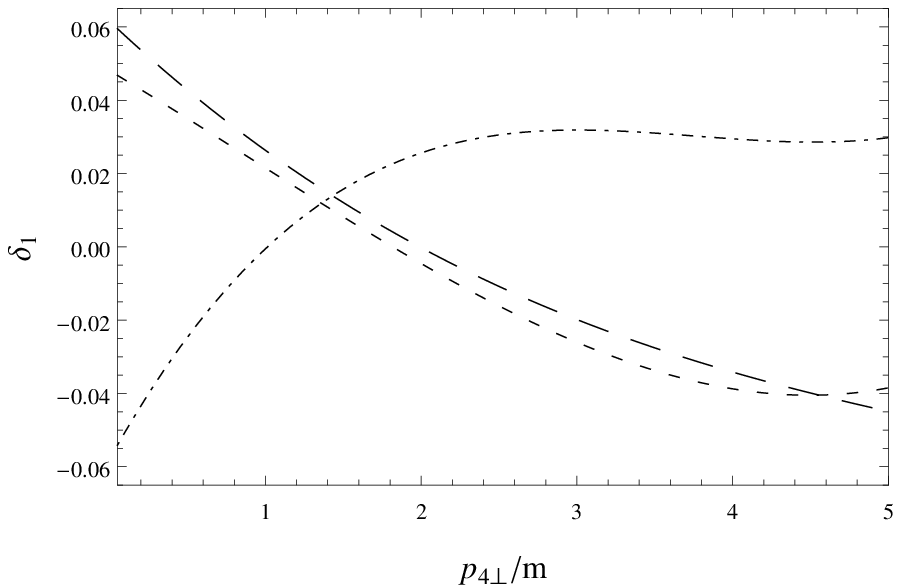}
	\caption{Left panel: the dependence of $d\sigma/dp_{4\perp}$ on $p_{4\perp}/m$ in units  $\sigma_0/m=\alpha^2\eta^2/m^3$ for  $Z=79$, $\varepsilon_1/m=100$; solid curve  is the exact result and dotted curve is the Born result. Right panel: the  quantity $\delta_1$ as the function of $p_{4\perp}/m$, where $\delta_1$ is  the deviation of the approximate result for $d\sigma/dp_{4\perp}$ from the exact one in units of the exact cross section. Dash-dotted curve is obtained without account for the interference between  $T$ and $\widetilde{T}$, dashed curve is obtained without contributions of $T^{(0)}$ and $\widetilde{T}^{(0)}$, and long-dashed curve corresponds to $\lambda=0$.}
\label{dif4}
\end{figure}

It follows from Fig.~\ref{dif2} that the deviation of the results obtained for $\lambda=1$ from that obtained for  $\lambda=0$ is
noticeable and  negative in the vicinity of the pick and small and positive in the wide region outside the pick. However, these two deviations (positive and negative) strongly compensate each other in the cross section integrated over both electron transverse momenta  $\bm p_{2\perp}$ and  $\bm p_{3\perp}$. This statement is illustrated in Fig.~\ref{dif4},  where
the cross sections   differential over the positron transverse momentum, $d\sigma/dp_{4\perp}$ is shown for $Z=79$ and $\varepsilon_1=100 m$.

Again, the Born result differs significantly from the exact one.  It is seen  that all relative deviations $\delta_1$  depicted in the right panel are noticeable. Then,
the results obtained for $\lambda=0$ and that without contributions  $T^{(0)}$ and $\widetilde{T}^{(0)}$ are very close to each other. This means that  account for the Coulomb corrections to scattering leads to a very small shift of the integrated cross section $d\sigma/dp_{4\perp}$, in contrast to the cross section $d\sigma/dp_{2\perp}$. Such suppression is similar to that found in our resent paper \cite{KM2017} at the consideration of $e^+e^-$ pair electroproduction by a heavy charged particle in the atomic field.

\begin{figure}[H]
	\centering
	\includegraphics[width=0.45\linewidth]{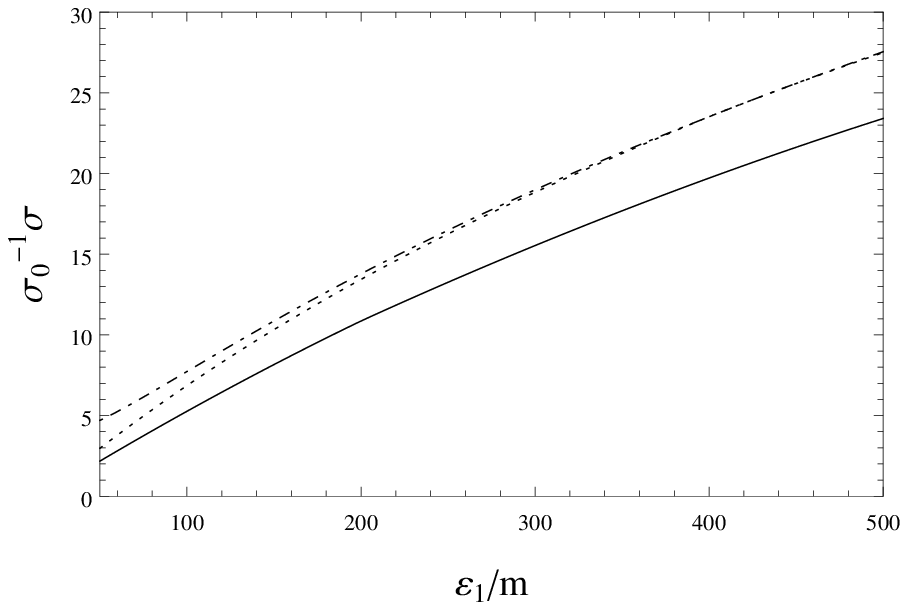}
\includegraphics[width=0.45\linewidth]{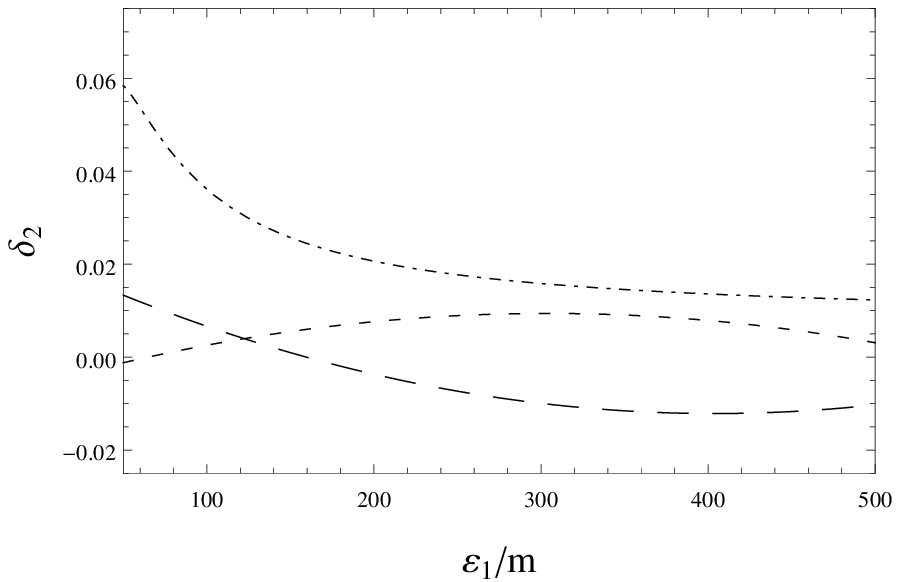}
		\caption{Left panel: the total cross section  $\sigma$ as the function of $\varepsilon_1/m$  in units  $\sigma_0=\alpha^2\eta^2/m^2$ for  $Z=79$; solid curve  is the exact result,  dotted curve is the Born result, and  dash-dotted curve is the ultra-relativistic asymptotics of the  Born result given by the formula  of Racah  \cite{Racah37}. Right panel: the  quantity $\delta_2$ as the function of $\varepsilon_1/m$, where $\delta_2$ is  the deviation of the approximate result for $\sigma$ from the exact one in units of the exact cross section. Dash-dotted curve is obtained without account for the interference between  $T$ and $\widetilde{T}$, dashed curve is obtained without contributions of $T^{(0)}$ and $\widetilde{T}^{(0)}$, and long-dashed curve corresponds to $\lambda=0$.}
\label{tot}
\end{figure}

At last, let us consider the total cross section $\sigma$ of the process under consideration. The cross section $\sigma$ for $Z=79$ as the function of  $\varepsilon_1/m$ is shown in the left panel in Fig.~\ref{tot}. In this picture solid curve  is the exact result, dotted curve is the Born result, and dash-dotted curve is the ultra-relativistic asymptotics of the  Born result given by the formula  of Racah  \cite{Racah37}. Note that a small deviation of our Born result at relatively small energies from the asymptotics of the Born result  is  due, first, to uncertainty of our result related to the uncertainty of low limit of integration over the energies of the produced particles, and secondly, to neglecting  identity of the final electrons in Ref.~\cite{Racah37}.

It is seen that the exact result  differs significantly from the Born one. In the right  panel of Fig.~\ref{tot} we show the relative deviation  $\delta_2$  of the approximate result for $\sigma$ from the exact one. Dash-dotted curve is obtained without account for the interference between  $T$ and $\widetilde{T}$, dashed curve is obtained without contributions  $T^{(0)}$ and $\widetilde{T}^{(0)}$, and long-dashed curve corresponds to $\lambda=0$. The corrections to the total cross section due to account for the  contributions  $T^{(0)}$ and $\widetilde{T}^{(0)}$, and  the Coulomb corrections to scattering  are small  even at moderate energy $\varepsilon_1$. The effect of the interference is more important at moderate energy and less important at high energies.

In our recent paper \cite{KM2016} the differential cross section of electroproduction by relativistic electron has been derived. For the differential cross section, we have pointed out that the Coulomb corrections to the scattering are the most noticeable  in the region $p_{2\perp}\sim \omega/\gamma$. On the basis of this statement, we have evaluated in the leading logarithmic approximation the Coulomb corrections to the total cross section, see Eq.~(33) of Ref.~\cite{KM2016}. However, as it is shown in the present paper, for the total cross section the  contribution of the  Coulomb corrections to  scattering  in the region  $p_{2\perp}\sim \omega/\gamma$ is  compensated strongly by the  contribution of the  Coulomb corrections to  scattering in the  wide region outside  $p_{2\perp}\sim \omega/\gamma$. As a result,  the Coulomb corrections to the total cross section derived in the leading logarithmic approximation does not affected by  account for the Coulomb corrections to  scattering.  This means that the coefficient in Eq.~(33) of Ref.~\cite{KM2016} should be  two times smaller and equal to that in the Coulomb corrections to the total cross section of $e^+e^-$ electroproduction by a relativistic heavy particle calculated in the leading logarithmic approximation. Note that an accuracy of the result obtained for the Coulomb corrections to the total cross section is very low because in electroproduction there is a strong compensation between the leading  and next-to-leading terms in the Coulomb corrections, see Ref.~\cite{LM2009}.

\section{Conclusion}
Performing  tabulations of the formula for the  differential cross section of  $e^+e^-$ pair electroproduction by a relativistic electron in the atomic field \cite{KM2016}, we have elucidated the importance of various contributions to the integrated cross sections of the process. It is shown that the Coulomb corrections are very important both for the differential cross section and for the integrated cross sections even for moderate values of the atomic charge number. This effect is mainly related to the Coulomb corrections to the amplitudes $T^{(1)}$ and  ${\widetilde T}^{(1)}$ due to the exact account of the interaction of the produced $e^+e^-$ pair with the atomic field (the Coulomb corrections to the amplitude of  $e^+e^-$ pair photoproduction by a virtual photon). There are also some other effects.   For the  cross section differential over the electron transverse  momentum, $d\sigma/dp_{2\perp}$, the account for the interference of the amplitudes  and the contribution of  virtual bremsstrahlung (the contribution of the amplitudes $T^{(0)}$ and  ${\widetilde T}^{(0)}$)
is noticeable.  The Coulomb corrections to  scattering is larger than these two effects but essentially smaller than the Coulomb corrections to the amplitude of  pair photoproduction by a virtual photon. However,  in the  cross section differential over the positron transverse  momentum, $d\sigma/dp_{4\perp}$, the  interference of the amplitudes  and the contribution of  virtual bremsstrahlung lead to the same corrections as the effect of the Coulomb corrections to scattering.  They are  of the same order as in the case of $d\sigma/dp_{2\perp}$. This means that there is a strong suppression of the effect of the Coulomb corrections to scattering in the cross section  $d\sigma/dp_{4\perp}$. Relative importance of various effects for the total cross section is the same as in the case of the  cross section $d\sigma/dp_{4\perp}$.

\section*{Acknowledgement}
This work has been supported by Russian Science Foundation (Project No. 14-50-00080). It has been also supported in part by
RFBR (Grant No. 16-02-00103).

\section*{Appendix}\label{app}
Here we present the explicit expression for the amplitude $T$, derived in  Ref.~\cite{KM2016}, with one modification. Namely, since we are going to investigate  the importance of the interaction of electrons with the momenta $\bm p_1$ and $\bm p_2$  with the atomic field, we introduce the parameter $\lambda$ which is equal to unity if this interaction is taken into account and equals to zero if one neglects this interaction. We write the amplitude $T$ in the form
$$T=T^{(0)}+T^{(1)}\,,\quad T^{(0)}=T^{(0)}_\parallel+T^{(0)}_\perp\,,\quad  T^{(1)}=T^{(1)}_\parallel+T^{(1)}_\perp\,,$$
where the helicity amplitudes $T^{(0)}_{\perp\parallel}$ read
\begin{align}\label{T0}
&T_\perp^{(0)}=\frac{8\pi A(\bm\Delta_0)}{\omega(m^2+\zeta^2)} \Big\{\delta_{\mu_1\mu_2}\delta_{\mu_3\bar\mu_4}
\Big[\frac{\varepsilon_3}{\omega^2}(\bm s_{\mu_3}^*\cdot \bm X)(\bm s_{\mu_3}\cdot\bm\zeta)(\varepsilon_1\delta_{\mu_1\mu_3}+\varepsilon_2\delta_{\mu_1\mu_4})\nonumber\\
&-\frac{\varepsilon_4}{\omega^2}(\bm s_{\mu_4}^*\cdot \bm X)(\bm s_{\mu_4}\cdot\bm\zeta) (\varepsilon_1\delta_{\mu_1\mu_4}+\varepsilon_2\delta_{\mu_1\mu_3})\Big]\nonumber\\
&-\frac{m\mu_1}{\sqrt{2}\varepsilon_1\varepsilon_2}R\delta_{\mu_1\bar\mu_2}\delta_{\mu_3\bar\mu_4}
(\bm s_{\mu_1}\cdot\bm\zeta)(-\varepsilon_3\delta_{\mu_1\mu_3}+\varepsilon_4\delta_{\mu_1\mu_4})\nonumber\\
&+\frac{m\mu_3}{\sqrt{2}\varepsilon_3\varepsilon_4}\delta_{\mu_1\mu_2}\delta_{\mu_3\mu_4}(\bm s_{\mu_3}^*\cdot\bm X)(\varepsilon_1\delta_{\mu_3\mu_1}+\varepsilon_2\delta_{\mu_3\bar\mu_1})
+\frac{m^2\omega^2}{2\varepsilon_1\varepsilon_2\varepsilon_3\varepsilon_4}R\delta_{\mu_1\bar\mu_2}\delta_{\mu_3\mu_4}\delta_{\mu_1\mu_3}\Big\}\,,\nonumber\\
&T_\parallel^{(0)}=-\frac{8\pi }{\omega^2}A(\bm\Delta_0)R\delta_{\mu_1\mu_2}\delta_{\mu_3\bar\mu_4}\,.
\end{align}
Here $\mu_i=\pm 1$ corresponds to the helicity of the particle with the momentum $\bm p_i$, $\bar\mu_i=-\mu_i$, and
\begin{align}\label{T0not}
&A(\bm\Delta)=-\frac{i\lambda}{\Delta_{\perp}^2}\int d\bm r\,\exp[-i\bm\Delta\cdot\bm r-i\chi(\rho)]\bm\Delta_{\perp}\cdot\bm\nabla_\perp V(r)\,,\nonumber\\
&\chi(\rho)=\lambda\int_{-\infty}^\infty dz\,V(\sqrt{z^2+\rho^2})\,,\quad\bm\rho=\bm r_\perp\,,\quad\bm\zeta=\frac{\varepsilon_3\varepsilon_4}{\omega}\bm\theta_{34}\,,\nonumber\\
&\omega=\varepsilon_3+\varepsilon_4\,, \quad \bm\Delta_{0\perp}=\varepsilon_2\bm\theta_{21}+\varepsilon_3\bm\theta_{31}+\varepsilon_4\bm\theta_{41}\,,\nonumber\\
&\Delta_{0\parallel}=-\frac{1}{2}\left[m^2\omega\left(\frac{1}{\varepsilon_1\varepsilon_2}+\frac{1}{\varepsilon_3\varepsilon_4}\right)+\frac{p_{2\perp}^2}{\varepsilon_2}+ \frac{p_{3\perp}^2}{\varepsilon_3}+\frac{p_{4\perp}^2}{\varepsilon_4}\right]\,,\nonumber\\
&R=\frac{1}{d_1d_2}[\Delta^2_{0\perp} (\varepsilon_1+\varepsilon_2)+2\varepsilon_1\varepsilon_2(\bm\theta_{12}\cdot\bm\Delta_{0\perp})]\,,\nonumber\\
&\bm X=\frac{1}{d_1}(\varepsilon_3\bm\theta_{23}+\varepsilon_4\bm\theta_{24})-\frac{1}{d_2}(\varepsilon_3\bm\theta_{13}+\varepsilon_4\bm\theta_{14})\,,\nonumber\\
&d_1=m^2\omega\varepsilon_1\left(\frac{1}{\varepsilon_1\varepsilon_2}+\frac{1}{\varepsilon_3\varepsilon_4}\right)+\varepsilon_2\varepsilon_3\theta_{23}^2
+\varepsilon_2\varepsilon_4\theta_{24}^2+\varepsilon_3\varepsilon_4\theta_{34}^2\,,\nonumber\\
&d_2=m^2\omega\varepsilon_2\left(\frac{1}{\varepsilon_1\varepsilon_2}+\frac{1}{\varepsilon_3\varepsilon_4}\right)+\varepsilon_2\varepsilon_3\theta_{31}^2
+\varepsilon_2\varepsilon_4\theta_{41}^2+(\varepsilon_3\bm\theta_{31}+\varepsilon_4\bm\theta_{41})^2\,,\nonumber\\
&\bm\theta_i=\bm p_{i\perp}/p_i \,,\quad \bm\theta_{ij}=\bm\theta_{i}-\bm\theta_{j}\,,
\end{align}
with $V(r)$ being the electron potential energy in the atomic field.    In the amplitude $T^{(0)}$ the interaction of the  produced $e^+e^-$ pair  with the atomic field is neglected, so that $T^{(0)}$  depends on the atomic potential in the same way as the  bremsstrahlung amplitude, see, e.g., Ref.~\cite{LMSS2005}.

The amplitudes $T^{(1)}_{\perp\parallel}$ have the following form
\begin{align}\label{T1C}
&T_\perp^{(1)}=\frac{8i\eta}{\omega \varepsilon_1}|\Gamma(1-i\eta)|^2 \int\frac{d\bm\Delta_\perp\, A(\bm\Delta_\perp+\bm p_{2\perp})F_a(Q^2)}{Q^2 M^2\,(m^2\omega^2/\varepsilon_1^2+\Delta_\perp^2)}\left(\frac{\xi_2}{\xi_1}\right)^{i\eta}
{\cal M}\,, \nonumber\\
&{\cal M}=-\frac{\delta_{\mu_1\mu_2}\delta_{\mu_3\bar\mu_4}}{\omega} \big[ \varepsilon_1(\varepsilon_3 \delta_{\mu_1\mu_3}-\varepsilon_4 \delta_{\mu_1\mu_4})
(\bm s_{\mu_1}^*\cdot\bm \Delta_\perp)(\bm s_{\mu_1}\cdot\bm I_1)\,\nonumber\\
&+\varepsilon_2(\varepsilon_3 \delta_{\mu_1\bar\mu_3}-\varepsilon_4 \delta_{\mu_1\bar\mu_4})(\bm s_{\mu_1}\cdot\bm \Delta_\perp)(\bm s_{\mu_1}^*\cdot\bm I_1)  \big]+\delta_{\mu_1\bar\mu_2}\delta_{\mu_3\bar\mu_4}\frac{m\omega\mu_1}{\sqrt{2}\varepsilon_1 }(\varepsilon_3 \delta_{\mu_1\mu_3}-\varepsilon_4 \delta_{\mu_1\mu_4})(\bm s_{\mu_1}
\cdot\bm I_1)\nonumber\\
&+\delta_{\mu_1\mu_2}\delta_{\mu_3\mu_4}\frac{m\mu_3}{\sqrt{2}}(\varepsilon_1 \delta_{\mu_1\mu_3}+\varepsilon_2 \delta_{\mu_1\bar\mu_3})(\bm s_{\mu_3}^*\cdot\bm \Delta_\perp)I_0
-\frac{m^2\omega^2}{2\varepsilon_1}\delta_{\mu_1\bar\mu_2}\delta_{\mu_3\mu_4}\delta_{\mu_1\mu_3}I_0\,,\nonumber\\
&T_\parallel^{(1)}=-\frac{8i\eta\varepsilon_3\varepsilon_4}{\omega^3}|\Gamma(1-i\eta)|^2 \int \frac{d\bm\Delta_\perp\, A(\bm\Delta_\perp+\bm p_{2\perp})F_a(Q^2)}{Q^2 M^2}\left(\frac{\xi_2}{\xi_1}\right)^{i\eta}\,I_0
\delta_{\mu_1\mu_2}\delta_{\mu_3\bar\mu_4}\,,
\end{align}
where $F_a(Q^2)$ is a atomic form factor,
and  the following notations are used
\begin{align}\label{T1Cnot}
&M^2=m^2\Big(1+\frac{\varepsilon_3\varepsilon_4}{\varepsilon_1\varepsilon_2}\Big)
+\frac{\varepsilon_1\varepsilon_3\varepsilon_4}{\varepsilon_2\omega^2} \Delta_\perp^2\,,\quad
\bm Q_\perp=\bm \Delta_\perp-\bm p_{3\perp}-\bm p_{4\perp}\,, \nonumber\\
&Q^2= Q_\perp^2+\Delta_{0\parallel}^2\,,\quad
\bm q_1=\frac{\varepsilon_3}{\omega}\bm \Delta_\perp- \bm p_{3\perp}\,,\quad \bm q_2=
\frac{\varepsilon_4}{\omega}\bm \Delta_\perp- \bm p_{4\perp} \,,\nonumber\\
&I_0=(\xi_1-\xi_2)F(x)+(\xi_1+\xi_2-1)(1-x)\frac{F'(x)}{i\eta}\,,\nonumber\\
&\bm I_1=(\xi_1\bm q_1+\xi_2\bm q_2)F(x)+(\xi_1\bm q_1-\xi_2\bm q_2)(1-x)\frac{F'(x)}{i\eta}\,,\nonumber\\
&\xi_1=\frac{M^2}{M^2+q_1^2}\,,\quad \xi_2=\frac{M^2}{M^2+q_2^2}\,,\quad x=1-\frac{Q_\perp^2\xi_1\xi_2}{M^2}\,,\nonumber\\
&F(x)=F(i\eta,-i\eta, 1,x)\,,\quad F'(x)=\frac{\partial}{\partial x}F(x)\,,\quad \eta=Z\alpha\,.
\end{align}
Note that the parameter $\lambda$ is contained solely  in the function $A(\bm\Delta)$, Eq.~\eqref{T0not}.

 \end{document}